\documentstyle[12pt]{article}
\def\be{\begin{equation}}\def\ee{\end{equation}}\def\l{\label}

\def\F{{\cal F}}

\def\L{{\cal L}} 
\def\M{{\cal M}}

\def\U{{\cal U}}

\makeatletter\ifcase\@ptsize\font\teneufm=eufm10 
\font\seveneufm=eufm7\font\fiveeufm=eufm5 
\font\teneusm=eusm10\font\seveneusm=eusm7 
\font\fiveeusm=eusm5\or\font\teneufm=eufm10 scaled 
\magstephalf\font\seveneufm=eufm7\font\fiveeufm=eufm5 
\font\teneusm=eusm10 scaled\magstephalf 
\font\seveneusm=eusm7\font\fiveeusm=eusm5\or 
\font\teneufm=eufm10 scaled\magstep1\font\seveneufm=eufm7 
\font\fiveeufm=eufm5\font\teneusm=eusm10 scaled\magstep1 
\font\seveneusm=eusm7\font\fiveeusm=eusm5\fi 
 
\newfam\eufmfam\newfam\eusmfam\textfont\eufmfam=\teneufm 
\scriptfont\eufmfam=\seveneufm 
\scriptscriptfont\eufmfam=\fiveeufm 
\textfont\eusmfam=\teneusm\scriptfont\eusmfam=\seveneusm 
\scriptscriptfont\eusmfam=\fiveeusm 
 
\def\frak{\ifmmode\let\next\frak@\else 
\def\next{\errmessage{Use\string\frak\space only in math mode}}\fi\next}\def\frak@#1{{\frak@@{#1}}} 
\def\frak@@#1{\fam\eufmfam#1} 
\def\sh{\ifmmode\let\next\sh@\else 
\def\next{\errmessage{Use\string\sh\space only in math mode}} 
\fi\next}\def\sh@#1{{\sh@@{#1}}} 
\def\sh@@#1{\fam\eusmfam#1} 
 
\ifcase\@ptsize\font\tenmsa=msam10\font\sevenmsa=msam7 
\font\fivemsa=msam5\font\tenmsb=msbm10 
\font\sevenmsb=msbm7\font\fivemsb=msbm5\or 
\font\tenmsa=msam10 scaled\magstephalf 
\font\sevenmsa=msam7\font\fivemsa=msam5 
\font\tenmsb=msbm10 scaled\magstephalf 
\font\sevenmsb=msbm7\font\fivemsb=msbm5\or 
\font\tenmsa=msam10 scaled\magstep1\font\sevenmsa=msam7 
\font\fivemsa=msam5\font\tenmsb=msbm10 scaled\magstep1 
\font\sevenmsb=msbm7\font\fivemsb=msbm5\fi 
 
\newfam\msafam\newfam\msbfam\textfont\msafam=\tenmsa 
\scriptfont\msafam=\sevenmsa 
\scriptscriptfont\msafam=\fivemsa\textfont\msbfam=\tenmsb 
\scriptfont\msbfam=\sevenmsb 
\scriptscriptfont\msbfam=\fivemsb 
 
\def\Bbb{\ifmmode\let\next\Bbb@\else 
\def\next{\errmessage{Use\string\Bbb\space only in math mode}} 
\fi\next}\def\Bbb@#1{{\Bbb@@{#1}}} 
\def\Bbb@@#1{\fam\msbfam#1}\def\hexnumber@#1{\ifnum#1<10 
\number#1\else\ifnum#1=10 A\else\ifnum#1=11 B\else\ifnum#1=12 C 
\else\ifnum#1=13 D\else\ifnum#1=14 E\else\ifnum#1=15 F\fi\fi\fi\fi\fi\fi\fi} 
\def\msa@{\hexnumber@\msafam}\def\msb@{\hexnumber@\msbfam} 
\mathchardef\square="0\msa@03 
 
\makeatother 
\newcommand{\HH}{{\Bbb H}}\newcommand{\RR}{{\Bbb R}} 
\newcommand{\CC}{{\Bbb C}}

\newcommand{\II}{{\Bbb I}} 
\newcommand{\TT}{{\Bbb T}} 
\newcommand{\ba}{\begin{array}} 
\newcommand{\ea}{\end{array}} 
\newcommand{\bea}{\begin{eqnarray}} 
\newcommand{\eea}{\end{eqnarray}}

\begin{document} 
 
\begin{titlepage} 
 
{\hfill MIT-CTP-2863} 
 
{\hfill DFPD98/TH/42} 
 
{\hfill US-FT/4-99} 
 
{\hfill hep-th/0003200} 
 
\begin{center}

\centerline{\Large\bf The Concept of a Noncommutative Riemann Surface}

\vspace{.999cm}{\centerline{\sc Gaetano BERTOLDI$^{1}$, 
Jos\'e M. ISIDRO$^{2}$, Marco MATONE$^{2}$ and Paolo PASTI$^{2}$}} 
\vspace{.2in} 
{\it $^{1}$ Center for Theoretical Physics\\ 
  Laboratory for Nuclear Science and Department of Physics\\ 
  Massachusetts Institute of Technology, Cambridge, MA 02139, USA\\ 
e-mail: bertoldi@ctp.mit.edu\\} 
\vspace{.08in} 
{\it $^{2}$ Dipartimento di Fisica ``G. Galilei'' -- Istituto Nazionale di 
Fisica Nucleare\\} 
 
{\it Universit\`a di Padova\\} 
 
{\it Via Marzolo, 8 -- 35131 Padova, Italy\\ 
e-mail: isidro, matone, pasti@pd.infn.it\\} 
\end{center} 
 
\centerline{\sc ABSTRACT} 
 
\vspace{0.6cm} 
 
\noindent 
We consider the compactification M(atrix) theory on a Riemann surface $\Sigma$ of genus 
$g>1$. A natural generalization of the case of the torus leads to construct a projective unitary 
representation of $\pi_1(\Sigma)$, realized on the Hilbert space of square integrable functions on 
the upper half--plane. A uniquely determined gauge connection, which in turn defines a gauged ${\rm 
sl}_2(\RR)$ algebra, provides the central extension. This has a geometric interpretation as the gauge 
length of a geodesic triangle, and corresponds to a 2--cocycle of the 2nd Hochschild cohomology group 
of the Fuchsian group uniformizing $\Sigma$. Our construction can be seen as a suitable 
double--scaling limit $N\to\infty$, $k\to-\infty$ of a $U(N)$ representation of $\pi_1(\Sigma)$, 
where $k$ is the degree of the associated holomorphic vector bundle, which can be seen as the 
higher--genus analog of 't Hooft's clock and shift matrices of QCD. We compare the above mentioned 
uniqueness of the connection with the one considered in the differential--geometric approach to the 
Narasimhan--Seshadri theorem provided by Donaldson. We then use our infinite dimensional 
representation to construct a $C^\star$--algebra which can be interpreted as a noncommutative 
Riemann surface $\Sigma_{\theta}$. Finally, we comment on the extension to higher genus of the 
concept of Morita equivalence.

\end{titlepage} 
 
\newpage 
 
\setcounter{footnote}{0} 
 
\renewcommand{\thefootnote}{\arabic{footnote}} 
\noindent 
{\it 1. The quotient conditions.} 
 
\noindent 
The $P_-=N/R$ sector of the discrete light--cone quantization  
of uncompactified M--theory is 
given by the supersymmetric quantum mechanics of $U(N)$ matrices. 
 In temporal gauge, the action reads 
\be 
S={1\over2R}\int dt\,{\rm Tr}\,\Big(\dot X^{\mu}\dot X_{\mu}+\sum_{\mu>\nu}[X^{ 
\mu},X^{\nu}]^2+i\Theta^T\dot\Theta-\Theta^T\Gamma_{\mu}[X^{\mu},\Theta]\Big), 
\l{lagrangiana}\ee 
where $\mu, \nu=1,\ldots, 9$. 
The compactification of M(atrix) theory \cite{MATRIX}--\cite{SUSS}  
as a model for M--theory 
\cite{REVTAYLOR} has been studied in \cite{TAYLOR}.  
In \cite{CDS}--\cite{CASALBUONI} it has been 
treated using noncommutative geometry \cite{CONNESETAL}.  
These investigations apply to the $d$--dimensional torus $T^d$,  
and have been further dealt with from 
various viewpoints in \cite{DOUGLAS}--\cite{IRAN}. These structures 
are also relevant in noncommutative string and gauge theories 
\cite{SWn,SchCorSch}. 
Let $e_{ij}$, $i,j=1,2$, generate a 2-dimensional 
lattice in $\RR^2$. In compactifying M(atrix) 
theory on the torus $\TT^2$ determined by this lattice one introduces unitary 
operators $\U_1$ and $\U_2$, defined on the covering space $\RR^2$ of $\TT^2$, 
such that 
\bea 
\U_i^{-1}X_j\U_i&=&X_j+2\pi e_{ij}, \quad i,j=1,2,\cr 
\U_i^{-1}X_a\U_i&=&X_a,\qquad\qquad\;\; a=3,\ldots,9\cr 
\U_i^{-1}\Theta \U_i&=&\Theta. 
\l{quotient}\eea 
By consistency the operators $\U_1$ and $\U_2$ commute, up to a 
constant phase: 
\begin{equation} 
{\cal U}_1\,{\cal U}_2=e^{2\pi i\theta}\,{\cal U}_2\, {\cal U}_1. 
\label{reltorus} 
\end{equation} 
In this paper we extend Eqs.(\ref{quotient})(\ref{reltorus}) 
to the case of compact Riemann surfaces of genus $g>1$. 
This is a first step towards the 
compactification of M(atrix) theory on a Riemann surface. 
The explicit solutions
and their supersymmetry properties will be considered
elsewhere.
 
A Riemann surface $\Sigma$ of genus $g>1$ is constructed  
as the quotient $\HH/\Gamma$, where 
$\HH$ is the upper half--plane, 
and $\Gamma\subset{\rm PSL}_2({\RR})$, 
$\Gamma\cong\pi_1(\Sigma)$, is a Fuchsian 
group acting on ${\HH}$ as 
\begin{equation}  
\gamma=\left(\begin{array}{c}a\\c 
\end{array}\begin{array}{cc}b\\d\end{array}\right)\in\Gamma,\;\qquad 
\gamma z={az+b\over cz+d}. 
\label{3} 
\end{equation} 
In the absence of elliptic and 
parabolic generators, the $2g$ Fuchsian generators $\gamma_j$  
satisfy 
\begin{equation} 
\prod_{j=1}^g\left(\gamma_{2j-1}\gamma_{2j}{\gamma_{2j-1}^{-1}} 
{\gamma_{2j}^{-1}}\right)={\II}. 
\label{3xpergamma} 
\end{equation} 
 
Inspired by M(atrix) theory, let us promote the complex coordinate  
$z =x+iy$ to an 
$N\times N$ complex matrix $Z=X+iY$, with $X=X^{\dagger}$ and $Y= Y^{\dagger}$. 
This would suggest defining fractional linear transformations of $Z$ through  
conjugation  
${\cal U} Z{\cal U}^{-1} =(aZ+b {\II})(cZ+d{\II})^{-1}$. 
However, taking the trace we see that 
this construction cannot be implemented for finite $N$. Thus we will consider some suitable modification. 
For the moment note that requiring the $\U_k$ to represent the 
$\gamma_k$ gives 
\begin{equation} 
\prod_{k=1}^g\left({\cal U}_{2k-1}\,{\cal U}_{2k}\,{\cal U}_{2k-1}^{-1}\, 
{\cal U}_{2k}^{-1}\right)=e^{2\pi i\theta}{\II}, 
\label{inonni} 
\end{equation}  
which generalizes the relation of the noncommutative torus (\ref{reltorus}). 
 
\vspace{.6cm} 
 
\noindent 
{\it 2. The noncommutative torus revisited.}

\noindent 
In order to compactify in higher genus it is necessary to extract 
some general guidelines from the case of the torus. 
In $g=1$ the fundamental group is Abelian. This implies that the associated differential generators 
commute, {\it i.e.} $[\partial_1,\partial_2]=0$, so it makes sense to apply the 
Baker--Campbell--Hausdorff (BCH) formula when computing the phase $e^{2\pi i\theta}$. 
On the contrary, the fundamental group of negatively curved Riemann surfaces is nonabelian, and 
the BCH formula is not useful. The derivation of the phase in $g=1$ by means of 
techniques alternative to the BCH formula will be the key point to solving the problem in $g>1$. 
 
Mimicking the case of $g=1$, one expects 
the building blocks for the solution to the quotient conditions in $g>1$ 
to have the form $e^{\L_n-\L_n^{\dagger}}$ or $e^{i(\L_n+\L_n^{\dagger})}$, 
for some {\it gauged ${\rm sl}_2(\RR)$ operators} $\L_n$ to be determined. 
We will show that finding such $\L_n$ is closely connected with 
the computation of the phase without using the BCH formula. 
In $g=1$ the BCH formula is useful, as the commutator between covariant 
derivatives can be a constant. On the contrary, in $g>1$, 
the $\L_n$ will be a sort of gauged ${\rm sl}_2(\RR)$ generators, 
and $[\L_n, \L_m]$ can never be a c--number. 
 
The solution to the quotient conditions in $g=1$ is expressed in terms of the exponential of 
covariant derivatives $\nabla_k$, $k=1,2$, so apparently we should use both $e^{\L_n-\L_n^{\dagger}}$ 
and $e^{i(\L_n+\L_n^{\dagger})}$ when passing to $g>1$. While the exponential 
$e^{\partial_k}$ will generate translations, the operator 
$e^{\L_n-\L_n^{\dagger}}$ will produce ${\rm PSL}_2(\RR)$ transformations. 
As the latter are real, we are forced to discard $e^{i(\L_n+\L_n^{\dagger})}$ and to use 
$e^{\L_n-\L_n^{\dagger}}$ only. This fact is strictly related to the nonabelian nature of the 
group $\pi_1(\Sigma)$. 
 
Let us consider the operators 
\be 
\U_k=e^{\lambda_k(\partial_k+iA_k)},\qquad \lambda_k\in\RR, \qquad k=1,2, 
\l{oiedj}\ee 
We also introduce the functions $F_k(x_1,x_2)$ defined by 
\be 
\U_k=F_ke^{\lambda_k\partial_k}F_k^{-1}, \qquad k=1,2. 
\l{osdjP}\ee 
The identity $Af(B)A^{-1}=f(ABA^{-1})$ and Eq.(\ref{osdjP}) give $(\partial_k+iA_k)F_k=0$. 
Also note that $e^{\lambda_k\partial_k}F_k^{-1}(\{x_k\})= 
F_k^{-1}(\{x_j+\delta_{jk}\lambda_k\}) e^{\lambda_k\partial_k}$. 
Therefore, defining $G_k(x_1, x_2)$ by 
\be 
\U_k=G_ke^{\lambda_k\partial_k}, \qquad k=1,2, 
\l{kouw}\ee 
we conclude that 
\be 
F_k(\{x_j+\delta_{jk}\lambda_k\})=G_k^{-1}(\{x_j\})F_k(\{x_j\}). 
\l{iwhqT}\ee 
 
The unitary operators $\U_k$ can be used to derive the phase 
of Eq.(\ref{reltorus}). First we note that pulling the derivatives 
to the right we get 
$$ 
\U_1\U_2\U_1^{-1}\U_2^{-1}= 
F_1e^{\lambda_1\partial_1}F_1^{-1}F_2e^{\lambda_2\partial_2}F_2^{-1} 
F_1e^{-\lambda_1\partial_1}F_1^{-1}F_2e^{-\lambda_2\partial_2}F_2^{-1} 
$$ 
$$ 
=F_1(x_1,x_2)F_1^{-1}(x_1+\lambda_1,x_2)F_2(x_1+\lambda_1,x_2) 
F_2^{-1}(x_1+\lambda_1,x_2+\lambda_2) 
$$ 
\be 
\times F_1(x_1+\lambda_1,x_2+\lambda_2) 
F_1^{-1}(x_1,x_2+\lambda_2)F_2(x_1,x_2+\lambda_2)F_2^{-1}(x_1,x_2). 
\l{iudwgd}\ee 
Let us consider the curvature of $A=A_1dx_1+A_2dx_2$ 
\be 
F=dA=(\partial_1A_2-\partial_2A_1)dx_1\wedge dx_2=F_{12}dx_1\wedge dx_2. 
\l{lidsq}\ee 
The constant--curvature connection is the unique possible choice 
to get a constant phase, so we set $F_{12}=2\pi\theta$. To be explicit we pick the gauge 
$ A_1=-{\pi\theta}x_2$, $A_2={\pi\theta}x_1$, so that 
$\U_1=e^{-i{\pi\lambda_1\theta}x_2}e^{\lambda_1\partial_1}$, 
$\U_2=e^{i{\pi\lambda_2\theta}x_1}e^{\lambda_2\partial_2}$, and 
\be 
G_1=e^{-i{\pi\lambda_1\theta}x_2}=e^{i\lambda_1A_1}, 
\qquad 
G_2=e^{i{\pi\lambda_2\theta}x_1}=e^{i\lambda_2A_2}, 
\l{jriuei}\ee 
so Eq.(\ref{iwhqT}) reads 
\be 
F_1(x_1+\lambda_1,x_2)=e^{i{\pi\lambda_1\theta}x_2}F_1(x_1,x_2), 
\qquad 
F_2(x_1,x_2+\lambda_2)=e^{-i{\pi\lambda_2\theta}x_1}F_2(x_1,x_2). 
\l{oicePo}\ee 
The solution is $F_1=e^{i{\pi\theta}x_1x_2}f_1(x_2)$,  
$F_2=e^{-i{\pi\theta}x_1x_2}f_2(x_1)$, 
with $f_1$ ($f_2$) an arbitrary function of $x_2$ ($x_1$). 
Substituting this into (\ref{iudwgd}) we get Eq.(\ref{reltorus}), as we would 
using BCH. 
 
{}From (\ref{kouw})(\ref{jriuei}) one would understand that the 
connection in (\ref{oiedj}) can be simply pulled to the left. However, this is the case only 
if one chooses a particular gauge, as in general we have $e^{\lambda_k(\partial_k+iA_k)}\ne 
e^{i\lambda_kA_k}e^{\lambda_k\partial_k}$. Indeed, under the gauge transformation 
$A_k\longrightarrow A_k+\partial_k\chi$, we have  
\be 
e^{\lambda_k(\partial_k+iA_k+i\partial_k\chi)}= 
e^{-i\chi}e^{\lambda_k(\partial_k+iA_k)}e^{i\chi}= 
e^{i\chi(\{x_j+\delta_{jk}\lambda_j\})-i\chi(\{x_j\})}e^{\lambda_k(\partial_k+iA_k)}, 
\l{oidsAZZOlK9}\ee 
whereas under a gauge transformation, $e^{i\lambda_kA_k}e^{\lambda_k\partial_k}$ 
is multiplied by $e^{i\lambda_k\chi(\{x_j\})}$. 
It is easily seen that the correct expression is 
\be 
e^{\lambda_k(\partial_k+iA_k)}=e^{i\int^{x_k+\lambda_k}_{x_k}da_kA_k}e^{\lambda_k\partial_k}, 
\l{iuxh0w}\ee 
where in the integrand one has $A_1(a_1,x_2)$ if $k=1$ and 
$A_2(x_1,a_2)$ if $k=2$. 
In (\ref{iuxh0w}) we used a shorthand notation; the integration limits 
should be written more precisely as 
$\int^{\{x_j+\delta_{jk}\lambda_j\}}_{\{x_j\}}$. In particular, 
the contour is easily recognized as the path joining $x_k$ and $x_k+\lambda_k$ along the 
line with $x_{j\ne k}$ fixed. Since on the torus we can choose the zero curvature metric, straight 
lines correspond to geodesics of the metric. Thus, the above contour is the geodesic joining 
$\{x_j\}$ with $\{x_j+\delta_{jk}\lambda_k\}$. A direct check of Eq.(\ref{iuxh0w}) is that  
\be 
e^{\lambda_k(\partial_k+iA_k)}=e^{-i\int_{x_k^0}^{x_k} da_k  
A_k}e^{\lambda_k\partial_k}e^{i\int_{x_k^0}^{x_k}da_kA_k}= 
e^{i\int^{x_k+\lambda_k}_{x_k}da_kA_k}e^{\lambda_k\partial_k}, 
\l{pdowjdJ}\ee 
where we used the property that $\partial_k\int_{x_{k0}}^{x_k}da_kA_k=A_k(x_1,x_2)$. 
This is a distinguished feature due to the flatness of the torus that does not 
hold in $g>1$. However, we redefine the contour integral for the 
torus in a way which easily generalizes to higher genus, namely, 
 
\noindent 
{\it The contour integral is along the geodesic, with respect to the constant curvature 
metric, joining the points with coordinates $\{x_j\}$ and $\{x_j+\delta_{jk}\lambda_k\}$}.

Due to the fact that along the integration contour either $dx_1=0$ or $dx_2=0$, we can replace 
$da_kA_k$ with $A$: 
\be 
\U_k=e^{\lambda_k(\partial_k+iA_k)}= e^{-i\int_{x_k^0}^{x_k}A} 
e^{\lambda_k\partial_k}e^{i\int_{x_k^0}^{x_k}A}= 
e^{i\int_{x_k}^{x_k+\lambda_k} 
A}e^{\lambda_k\partial_k}, 
\l{thereasonforthis}\ee 
so that $F_k=e^{-i\int_{x_k^0}^{x_k}A}$. Even if on the 
torus the $F_k$ are not essential, we introduced them as their higher--genus analog 
will lead to a new class of functions. By Stokes' theorem 
$$ 
\U_1\U_2\U_1^{-1}\U_2^{-1}= 
$$ 
$$ 
\exp\left[i\int_{(x_1,x_2)}^{(x_1+\lambda_1,x_2)}A+ 
i\int_{(x_1+\lambda_1,x_2)}^{(x_1+\lambda_1,x_2+\lambda_2)}A 
+i\int_{(x_1+\lambda_1,x_2+\lambda_2)}^{(x_1,x_2+\lambda_2)}A 
+i\int_{(x_1,x_2+\lambda_2)}^{(x_1,x_2)}A\right] 
$$ 
\be 
=\exp\left(i\oint_{\partial\F}A\right)=\exp\left(i\int_\F F\right) 
=e^{2\pi i\lambda_1\lambda_2 \theta}, 
\l{opijwJD}\ee 
where $\F$ is a fundamental domain for the torus. 
Note that $\lambda_1\lambda_2$ is the area of the torus. Normalizing the area to $1$, 
we get Eq.(\ref{reltorus}). 
 
We now show that the only possible 
connection leading to a constant value of $\oint_{\partial\F}A$ 
is the one with constant curvature. In order to denote the dependence 
on the basepoint of the domain we use the notation $\F_{x_1x_2}$. 
Independence from $(x_1,x_2)$ means that 
\be 
\int_{\F_{x_1x_2}}F =\int_{\F_{x_1'x_2'}}F, 
\l{fjewhft}\ee 
for any $(x_1',x_2')\in\RR^2$. Any point in $\RR^2$ can be obtained by 
a translation $(x_1,x_2)\to(x_1',x_2')=\mu(x_1,x_2)\equiv (x_1+b_1,x_2+b_2)$. 
Noticing that 
$\F_{x_1'x_2'}=\mu\F_{x_1x_2}$, 
we see that Eq.(\ref{fjewhft}) is satisfied only if the 
curvature two--form $F$ is invariant under arbitrary translations 
of $(x_1,x_2)$. This fixes $F$ to be a constant 
two--form. 
 
The above investigation captures the essence of the construction in $g=1$, 
somehow extracting it from its specific context. This is 
very useful to reformulate the problem of deriving 
a projective unitary representation of the fundamental group of a class of 
manifolds which is much more general than 
the torus. We can say that in order to get 
a projective unitary representation of the fundamental group of 
a given manifold $\M$ by means of operators acting on 
the space $L^2(\M)$, we should consider the previous well--defined 
guidelines. 
 
\vspace{.6cm} 
 
\noindent 
{\it 3. Projective unitary representation of $\pi_1(\Sigma)$ on $L^2(\HH)$.} 
 
\noindent 
We now apply the above general 
analysis to the case of higher genus Riemann surfaces. 
We start by first considering a unitary representation of $\pi_1(\Sigma)$ 
realized on $L^2(\HH)$ (the analog of $e^{\lambda_k\partial_k}$). 
For $n=-1,0,1$ and 
$e_n(z)=z^{n+1}$ let us set $\ell_n=e_n(z)\partial_z$. Define 
\begin{equation}  
L_n=e_n^{-1/2}\ell_ne^{1/2}_n= e_n\left(\partial_z+{1\over2} 
{e'_n\over e_n}\right). 
\label{glielleenne} 
\end{equation}  
They satisfy the ${\rm sl}_2(\RR)$ algebra 
\begin{equation} 
[L_m,L_n]=(n-m)L_{m+n},\quad [\bar L_m,L_n]=0, 
\quad 
[L_n,f]=z^{n+1}\partial_zf. 
\label{x9ewixh} 
\end{equation} 
For $k=1,2,\ldots, 2g$, consider the operators 
\begin{equation} 
T_k=e^{\lambda_{-1}^{(k)}(L_{-1}+\bar L_{-1})}e^{\lambda_0^{(k)}(L_0+\bar L_0)} 
e^{\lambda_{1}^{(k)}(L_{1}+\bar L_{1})}, 
\label{iVcappa} 
\end{equation} 
with the $\lambda_n^{(k)}$ picked such that 
\be 
T_kzT_k^{-1}=\gamma_kz={a_kz+b_k\over  
c_kz+d_k}, 
\l{caaasss}\ee 
so that by (\ref{3xpergamma})  
\begin{equation} 
\prod_{k=1}^g\left( T_{2k-1}T_{2k}T_{2k-1}^{-1}T_{2k}^{-1} 
\right)={\II}. 
\label{thetees} 
\end{equation}  
On $L^2(\HH)$ we have the scalar product $\langle\phi|\psi\rangle=\int_\HH d\nu\bar\phi\psi$, 
with $d\nu(z)=idz\wedge d\bar z/2=dx\wedge dy$. One can check that the $T_k$ provide a unitary 
representation of $\Gamma$. 
 
For any function $F$ satisfying $|F|=1$, we define the operators 
\be 
{\cal L}_n^{(F)}=F(z,\bar z) L_n F^{-1}(z,\bar z) 
=e_n\left(\partial_z+{1\over 2}{e'_n\over e_n}- 
\partial_z \ln F(z,\bar z)\right), 
\label{glielleennecal} 
\end{equation}  
which also satisfy the algebra (\ref{x9ewixh}). 
Its adjoint is given by 
\begin{equation} 
{\cal L}_n^{(F)\dagger}=-F\overline{e^{1/2}_n}\partial_{\bar z}\overline {e^{1/2}_n}F^{-1} 
=-\bar{\cal L}_n^{(F^{-1})}. 
\label{abbastanzabasilare} 
\end{equation} 
We now observe that the operators 
\begin{equation} 
\Lambda_n^{(F)}= 
{\cal L}_n^{(F)}-{\cal L}_n^{(F)\dagger}={\cal L}_n^{(F)}+  
\bar{\cal L}_n^{(F^{-1})}, 
\label{zx4} 
\end{equation}  
enjoy the fundamental property that both their chiral  
components are gauged in the same way by the function $F$, that is 
\begin{equation} 
\Lambda_n^{(F)}=F(L_n+\bar L_n)F^{-1}, 
\label{zx5} 
\end{equation}  
while also satisfying the ${\rm sl}_2(\RR)$ algebra: 
\begin{equation} 
[\Lambda_m^{(F)},\Lambda_n^{(F)}]=(n-m)\Lambda_{m+n}^{(F)}, 
\qquad 
[\Lambda_n^{(F)},f]=(z^{n+1}\partial_z+{\bar z}^{n+1}\partial_{\bar z})f. 
\label{soddisfanosoddisfano} 
\end{equation} 
Furthermore, since $\Lambda_n^{(F)\dagger}=-\Lambda_n^{(F)}$, the operators 
$e^{\Lambda_n^{(F)}}=Fe^{L_n+\bar L_n}F^{-1}$ are unitary. 
 
Let $b$ be a real number, and $A$ a Hermitean connection  
to be identified presently. Set 
\begin{equation}  
{\cal U}_k=e^{ib\int_z^{\gamma_k z}A}T_k, 
\label{loperatoree} 
\end{equation} 
where the integration contour is taken to be the Poincar\'e geodesic connecting  
$z$ and $\gamma_k z$. 
As the gauging functions introduced in (\ref{glielleennecal}) we will  
take the 
$F_k(z,\bar z)$ solutions of the equation $F_kT_kF^{-1}_k=e^{ib\int_z^{\gamma_kz}A}T_k$,  
that is 
\begin{equation} 
F_k(\gamma_kz, \gamma_k\bar z)=e^{-ib\int_z^{\gamma_k z} A}F_k(z,\bar z). 
\label{thatis} 
\end{equation} 
With the choice (\ref{thatis}) for $F_k$, (\ref{zx5}) becomes 
\be 
\Lambda_{n,k}^{(F)}=F_k(L_n+\bar L_n)F^{-1}_k 
=z^{n+1}\left(\partial_z+{n+1\over2z}-\partial_z \ln F_k\right) 
+{\bar z}^{n+1}\left(\partial_{\bar z}+{n+1\over2{\bar z}}- 
\partial_{\bar z} \ln 
F_k\right). 
\label{zx5bisse} 
\ee 
The $\Lambda_{n,k}^{(F)}$ satisfy the algebra 
$$ 
[\Lambda_{m,j}^{(F)},\Lambda_{n,k}^{(F)}] 
=(n-m)\Lambda_{m+n,j}^{(F)} 
+F_k^{-1}| e_n|\Lambda_{n,k}^{(F)}|e_n|^{-1}F_k 
F_j^{-1}|e_m|\Lambda_{m,j}^{(F)}|e_m|^{-1}F_j(\ln F_j- \ln F_k), 
$$ 
\be 
[\Lambda_{n,k}^{(F)},f]=(z^{n+1}\partial_z+{\bar z}^{n+1}\partial_{\bar z})f. 
\label{commutator} 
\end{equation} 
Upon exponentiating $\Lambda_{n,k}^{(F)}$ one finds 
\begin{equation} 
{\cal U}_k=e^{\lambda_{-1}^{(k)}\Lambda_{-1,k}^{(F)}}\,e^{\lambda_0^{ 
(k)}\Lambda_{0,k}^{(F)}}\,e^{\lambda_{1}^{(k)}\Lambda_{1,k}^{(F)}}, 
\label{zx11} 
\end{equation} 
that is, the ${\cal U}_k$ are unitary, and 
\begin{equation} 
{\cal U}_k^{-1}=T_k^{-1}e^{-ib\int_z^{\gamma_kz}A}= 
e^{-ib\int_{\gamma_k^{-1}z}^zA}T_k^{-1}. 
\label{iosdq899999} 
\end{equation} 
It is immediate to see that the ${\cal U}_k$ defined in (\ref{loperatoree})  
satisfy (\ref{inonni}) for a certain value of $\theta$:\footnote{The differential representation of ${\rm 
PSL}_2({\RR})$ acts in reverse order with respect to the one by matrices.} 
$$ 
\prod_{k=1}^g\left({\cal U}_{2k-1}{\cal U}_{2k}{\cal U}_{2k-1}^{\dagger} 
{\cal U}_{2k}^{\dagger}\right)  
=e^{ib\int_z^{\gamma_1z}A}T_1e^{ib\int_z^{\gamma_2z}A}T_2 
e^{-ib\int_{\gamma_1^{-1}z}^zA}T_1^{-1}e^{-ib\int_{\gamma_2^{-1}z}^zA} 
T_2^{-1}\ldots 
$$ 
$$ 
=\exp\left[ib\left(\int_z^{\gamma_1z}+\int_{\gamma_1z}^{\gamma_2\gamma_1z} 
+\int_{\gamma_2\gamma_1z}^{\gamma_1^{-1}\gamma_2\gamma_1z} 
+\int_{\gamma^{-1}_1 
\gamma_2\gamma_1z}^{\gamma_2^{-1}\gamma_1^{-1}\gamma_2\gamma_1z}+ 
\ldots\right)A\right] 
\prod_{k=1}^g\left(T_{2k-1} 
T_{2k}T_{2k-1}^{-1}T_{2k}^{-1}\right) 
$$ 
\be 
=e^{ib\oint_{\partial{\cal F}_z}A}, 
\label{abbiamox1} 
\end{equation}  
where ${\cal F}_z=\{z,\gamma_1 z,\gamma_2\gamma_1z,\gamma_1^{-1} 
\gamma_2\gamma_1 z,\ldots\}$ 
is a fundamental domain for $\Gamma$. 
The basepoint $z$, plus the action of the Fuchsian generators on it,  
determine 
${\cal F}_z$, as the vertices are joined by geodesics.  
 
For (\ref{abbiamox1}) to provide a projective unitary representation 
of $\Gamma$, $\int_{{\cal F}_z} dA$ should be $z$--independent. 
Changing $z$ to 
$z'$ can be expressed as $z\to z'=\mu z$ for some  
$\mu\in{\rm PSL}_2({\RR})$. Then 
${\cal F}_z\rightarrow {\cal F}_{\mu z}= 
\{\mu z,\gamma_1\mu z,\gamma_2\gamma_1 \mu 
z,\gamma_1^{-1}\gamma_2 \gamma_1 
\mu z,\ldots\}$. 
Now consider 
${\cal F}_z\rightarrow \mu 
{\cal F}_z= 
\{\mu z,\mu\gamma_1 z,\mu\gamma_2\gamma_1 z, 
\mu\gamma_1^{-1}\gamma_2 \gamma_1 z,\ldots\}$. 
The congruence $\mu {\cal F}_z\cong{\cal F}_{\mu z}$ follows from  
two facts: that the 
vertices are joined by geodesics, and that ${\rm PSL}_2({\RR})$ maps geodesics 
into geodesics. Since $\Gamma$ is defined up to conjugation,  
$\Gamma\to\mu\Gamma\mu^{-1}$, if $\mu{\cal F}_z$ is a fundamental domain,  
so is ${\cal F}_{\mu z}$. Thus, 
to have $z$--independence we need $\forall\mu\in{\rm PSL}_2({\RR})$ 
\begin{equation} 
\int_{{\cal F}_z}dA=\int_{{\cal F}_{\mu z}}dA=\int_{\mu{\cal F}_z}dA 
=\int_{{\cal F}}dA. 
\label{0ijqI} 
\end{equation} 
This fixes the 
(1,1)--form $dA$ to be ${\rm PSL}_2({\RR})$--invariant. 
It is well known that the Poincar\'e form is the unique 
${\rm PSL}_2({\RR})$--invariant (1,1)--form, up to an overall constant 
factor. This is a particular case of a more general fact \cite{NOI}. 
The Poincar\'e metric $ds^2=y^{-2}|dz|^2=2g_{z\bar z}|dz|^2= 
e^{\varphi}|dz|^2$ has curvature 
$R=-g^{z\bar z}\partial_z\partial_{\bar z}\ln \, g_{z\bar z}=-1$, so that  
$\int_{{\cal F}} d\nu e^{\varphi}= -2\pi\chi(\Sigma)$,  
where $\chi(\Sigma)=2-2g$ is the Euler 
characteristic. As the Poincar\'e (1,1)--form is  
$dA=e^\varphi d\nu$, 
this uniquely determines the gauge field to be  
\begin{equation}   
A=A_zdz+A_{\bar z}d\bar z={dx\over y}, 
\label{Aconn} 
\end{equation}   
modulo gauge transformations. Using  
$\oint_{\partial{\cal F}}A=\int_{\cal F} dA$ 
 we finally have that (\ref{abbiamox1}) becomes 
\begin{equation} 
\prod_{k=1}^g\left({\cal U}_{2k-1}{\cal U}_{2k}{\cal U}_{2k-1}^{\dagger} 
{\cal U}_{2k}^{\dagger}\right)= 
e^{2\pi ib\chi(\Sigma)}. 
\label{perfect} 
\end{equation}

\vspace{.6cm} 
 
\noindent 
{\it 4. Nonabelian gauge fields.} 
 
\noindent 
 Up to now we considered the case in which the connection is 
Abelian. However, it is easy to extend our construction to the 
nonabelian case in which the gauge group 
$U(1)$ is replaced by 
$U(N)$. The operators ${\cal U}_k$ now become 
\begin{equation} 
{\cal U}_k=P e^{ib\int^{\gamma_k z}_z A} T_k, 
\label{cdo3jn}\end{equation} 
where the $T_k$ are the same as before, times the $N\times N$ identity matrix. 
Eq.(\ref{abbiamox1}) is replaced by 
\begin{equation} 
\prod_{k=1}^g\left({\cal U}_{2k-1}{\cal U}_{2k}{\cal U}_{2k-1}^{\dagger} 
{\cal U}_{2k}^{\dagger}\right) 
= P e^{ib\oint_{\partial{\cal F}_z}A}. 
\label{abbiamox1nonabeliana} 
\end{equation} 
Given an integral 
along a closed contour $\sigma_z$ with basepoint 
$z$, the path--ordered exponentials for 
a connection $A$ and its gauge transform 
$A^U=U^{-1}AU+U^{-1}dU$ are related by \cite{THOMPSON} 
\begin{equation} 
P e^{i\oint_{\sigma_z}A}=U(z) 
P e^{i\oint_{\sigma_z}A^U}U^{-1}(z)= 
U(z)P e^{i\oint_{\sigma_z}d\sigma^\mu\int_0^1ds s\sigma^\nu 
U^{-1}(s\sigma)F_{\nu\mu} 
(s\sigma)U(s\sigma)}U^{-1}(z). 
\label{buonina}\end{equation}  
Applying this to (\ref{abbiamox1nonabeliana}), we see that the only 
possibility to get a coordinate--independent phase is for 
the curvature (1,1)--form $F=dA+[A,A]/2$ 
to be the identity matrix in the gauge indices times a (1,1)--form 
$\eta$, that is $F=\eta{\II}$. 
It follows that 
\begin{equation} 
P e^{ib\oint_{\partial{\cal F}}A}= 
e^{ib\int_{{\cal F}}F}. 
\label{boh}\end{equation} 
This is only a necessary 
condition for coordinate--independence. However, this is the same 
as the Abelian case so that $\eta$ should be 
proportional to the Poincar\'e (1,1)--form. 
 
Denoting by $E$ the vector bundle on which $A$ is defined,  
we have $k={\rm deg}\,(E)={1\over 2\pi}{\rm tr}\,\int_{{\cal F}} F$. 
Set $\mu(E)=k/N$ so that 
$\int_{{\cal F}}F=2\pi \mu(E){\II}$ and 
$\eta=-{\mu(E)\over \chi(\Sigma)} e^\varphi d\nu$, {\it i.e.} 
\begin{equation} 
F=2\pi\mu(E)\omega {\II}, 
\label{zummmolloo}\end{equation} 
where $\omega=\left(e^\varphi/\int_{{\cal F}}d\nu e^\varphi\right) 
d\nu$. Thus, by 
(\ref{boh}) we have that Eq.(\ref{abbiamox1nonabeliana}) becomes 
\begin{equation} 
\prod_{k=1}^g\left({\cal U}_{2k-1}{\cal U}_{2k}{\cal U}_{2k-1}^{\dagger} 
{\cal U}_{2k}^{\dagger}\right) 
= e^{2\pi i b\mu(E)}{\II}, 
\label{abbiamox1nonabelianaduino} 
\end{equation} 
which provides a projective unitary representation 
of $\pi_1(\Sigma)$ on $L^2(\HH,\CC^N)$. 
 
\vspace{.6cm} 
 
\noindent 
{\it 5. Hochschild cohomology and gauge lengths.} 
 
\noindent 
 A basic object is the 
{\it gauge length} function $d_{A}(z,w)=\int^w_zA$, where the contour 
integral is along the Poincar\'e geodesic connecting $z$ and $w$. In the Abelian case  
\begin{equation} 
d_{A}(z,w)=\int^{{\rm Re}\,w}_{{\rm Re}\,z}{dx\over y} 
=-i\ln\left( {z-\bar w\over w-\bar z}\right), 
\label{logaritmo} 
\end{equation} 
which is equal to the angle $\alpha_{zw}$ spanned by the arc of geodesic 
connecting $z$ and $w$. Observe that the gauge length of the geodesic connecting 
two punctures, {\it i.e.} two points on the real line, is $\pi$. This is to be compared  
with the usual divergence of the Poincar\'e distance. Under a ${\rm PSL}_2({\RR})$--transformation 
$\mu$, we have ($\mu_x\equiv\partial_x \mu x$) 
\begin{equation} 
d_{A}\left(\mu z,\mu w\right)=d_{A}(z,w)  
-{i\over 2}\ln\left({\mu_z\bar \mu_w\over\bar\mu_z \mu_w}\right). 
\label{ioeuIpo} 
\end{equation} 
Therefore, the gauge length of an $n$--gon 
\begin{equation}  
d^{(n)}_{A}(\{z_k\})= \sum_{k=1}^nd_{A}(z_k,z_{k+1}) 
=\pi(n-2)-\sum_{k=1}^n\alpha_k, 
\label{poli4w} 
\end{equation} 
where $z_{n+1}\equiv z_1$, $n\geq 3$, and $\alpha_k$ are the internal angles, 
is ${\rm PSL}_2({\RR})$--invariant. 
 
We now show that the length of the triangle 
is proportional to the Hochschild 2--cocycle of $\Gamma$. 
The Fuchsian generators $\gamma_k\in\Gamma$ are projectively represented  
by means of unitary operators ${\cal U}_k$ acting on $L^2({\HH})$. The 
product $\gamma_k\gamma_j$ is represented by 
${\cal U}_{jk}$, which equals ${\cal U}_j{\cal U}_k$ up to a phase: 
\begin{equation} 
{\cal U}_j{\cal U}_k=e^{2\pi i\theta(j,k)}{\cal U}_{jk}. 
\label{projectivity} 
\end{equation} 
Associativity implies 
\begin{equation} 
\theta(j,k) + \theta(jk,l)= 
\theta(j,kl) + \theta(k,l). 
\label{cocycle} 
\end{equation} 
We can easily determine $\theta(j,k)$: 
\be 
{\cal U}_j{\cal U}_k=\exp\left(ib\int_z^{\gamma_j 
z}A+ib\int_{\gamma_jz}^{\gamma_k\gamma_jz}A 
-ib\int_z^{\gamma_k\gamma_jz}A\right){\cal U}_{jk} = 
\exp\left(ib\int_{\tau_{jk}}A\right){\cal U}_{jk}, 
\label{coc} 
\end{equation} 
where $\tau_{jk}$ denotes the geodesic triangle with vertices $z$,  
$\gamma_jz$ and $\gamma_k\gamma_jz$. This identifies $\theta(j,k)$ as the gauge 
length of the perimeter of the geodesic triangle $\tau_{jk}$ times 
$b/2\pi$. By Stokes' theorem this is 
the Poincar\'e area of the triangle. One can check that $\theta(j,k)$ in fact 
satisfies (\ref{cocycle}). This phase has been considered 
in different contexts, such as the quantum Hall effect on $\HH$ \cite{CHMM} and 
Berezin's quantization of ${\HH}$ and Von Neumann algebras \cite{RADULESCU}. 
 
The information on the compactification of M(atrix) theory is encoded in 
the action of $\Gamma$ on ${\HH}$, plus a projective representation of $\Gamma$.  
The latter amounts to the choice of a phase. Physically inequivalent choices of 
$\theta(j,k)$ turn out to be in one--to--one correspondence with elements in the 2nd Hochschild 
cohomology group of $\Gamma$, which is $U(1)$. Hence $\theta=b\chi(\Sigma)$ is the unique 
parameter for this compactification ($\theta=b\mu(E)$ in the general case). 
 
The Poincar\'e metric is ${\rm PSL}_2(\RR)$ invariant whereas $A$ is not. So the equality 
$\oint_{\partial{\cal F}}A=\int_{\cal F}F$ should be a consequence of the fact that 
the variation of $A$ under a ${\rm PSL}_2(\RR)$ transformation, $z\to \mu z=(az+b)/(cz+d)$, 
corresponds to a total derivative. In fact we have 
\be 
{\rm PSL}_2(\RR):A\longrightarrow i{d\mu z+d\mu\bar z\over\mu z-\mu\bar z}= 
A-i\partial_z\ln(cz+d)dz+i\partial_{\bar z}\ln(c\bar z+d)d\bar z. 
\l{pokwI}\ee 
Since $cz+d$ has no zeroes, we have that 
$\ln(cz+d)$ is a genuine function on $\HH$. It follows that 
$-i\partial_z\ln(cz+d)dz+i\partial_{\bar z}\ln(c\bar z+d)d\bar z$, can be written as an external 
derivative so that Eq.(\ref{pokwI}) becomes 
\be 
{\rm PSL}_2(\RR):A\longrightarrow A+d \ln(\mu_z/\bar\mu_z)^{i\over 2}, 
\l{owichwP}\ee 
where $\mu_z\equiv\partial_z\mu z$. So a ${\rm PSL}_2(\RR)$--transformation of $A$ is equivalent to a gauge transformation. Under $A\to A+d\chi$ we have 
$\int_z^w A\longrightarrow\int_z^wA+\chi(w)-\chi(z)$, 
which for $\chi(z)=\ln(\mu_z/\bar\mu_z)^{i\over 2}$, becomes 
\be 
\int_z^w A\longrightarrow\int_z^wA+{i\over 2} 
\ln{\bar\mu_z\mu_w\over\mu_z\bar\mu_w}. 
\l{qsottonagaugetra2}\ee 
 
\vspace{.6cm} 
 
\noindent 
{\it 6. Preautomorphic forms.} 
 
\noindent 
 Another reason why the gauge--length function is important is that it 
also appears in the definition (\ref{thatis}) of the $F_k$. The latter functions, which apparently 
never appeared in the literature before, are of particular interest. By (\ref{thatis}) and 
(\ref{logaritmo}), 
\begin{equation} 
F_k(\gamma_k z,\gamma_k\bar z)=\left({\gamma_k z-\bar z\over  
z-\gamma_k \bar z}\right)^b\,F_k(z,\bar z). 
\label{effecapppa4} 
\end{equation} 
Since under a ${\rm PSL}_2(\RR)$ transformation the factor 
$(w-\bar z)/(z-\bar w)$ gets transformed by a factor 
which is typical of automorphic forms, we 
call the $F_k$ {\it preautomorphic forms}. 
Eq.(\ref{thatis}) indicates that finding the most general solution to  
(\ref{effecapppa4}) is 
a problem in geodesic analysis. In the case of the inversion 
$\gamma_kz=-1/z$ and $b$ an even integer, a solution to  
(\ref{effecapppa4}) is  
$F_k=\left(z/\bar z\right)^{b\over2}$. By (\ref{logaritmo}) 
$F_k=\left(z/ \bar z\right)^{b\over2}$ is related to the {\it $A$--length} of 
the geodesic connecting $z$ and $0$: 
\begin{equation}  
e^{{i\over2}b\int_z^0A}=F_k(z,\bar z)=\left({z\over\bar z}\right)^{b\over2}. 
\label{idwjhyu} 
\end{equation} 
An interesting formal solution to (\ref{effecapppa4}) is 
\begin{equation} 
F_k(z,\bar z)=\prod_{j=0}^\infty\left({\gamma_k^{-j}z-\gamma_k^{-j-1}  
\bar z\over 
\gamma_k^{-j-1} z-\gamma_k^{-j} \bar z}\right)^b. 
\label{pro} 
\end{equation} 
Consider the uniformizing map 
$J_{\HH}:{\HH}\longrightarrow\Sigma$, 
which enjoys the property $J_{\HH}(\gamma z)=J_{\HH}(z)$, 
$\forall\gamma\in\Gamma$. Then another solution to (\ref{effecapppa4}) is given by 
$G(J_{\HH},\bar J_{\HH})F_k$, where $G$ is an arbitrary function of the uniformizing map. 
We should require $|G|=1$ for $|F_k|=1$. 
 
\vspace{.6cm} 
 
\noindent 
{\it 7. Relation with Donaldson's approach to stable bundles.} 
 
\noindent 
 We now present some facts about 
projective, unitary representations of 
$\Gamma$ and the theory of holomorphic vector bundles 
\cite{KOBAYASHI,AtiyahSenGuptaFine}. 
Let $E\rightarrow \Sigma$  
be a holomorphic vector bundle over $\Sigma$ of rank $N$ and degree $k$. 
The bundle $E$ is called {\it stable}   
if the inequality $\mu(E')<\mu(E)$  
holds for every proper holomorphic subbundle $E'\subset E$.  
We may take $-N<k\leq 0$. We will further assume that  
$\Gamma$ contains a unique primitive elliptic  
element $\gamma_0$ of order $N$ ($i.e.$, $\gamma_0^N={\II}$),  
with fixed point $z_0\in{\HH}$ that projects to $x_0\in\Sigma$. 
 
Given the branching order $N$ of $\gamma_0$,  
let $\rho:\Gamma\to U(N)$ be an irreducible unitary representation. 
It is said {\it admissible} if  
$\rho(\gamma_0)=e^{-2\pi i k/N}{\II}$. 
Putting the elliptic element on the right--hand side, and 
setting $\rho_k\equiv\rho(\gamma_k)$, 
(\ref{3xpergamma}) becomes 
\begin{equation}  
\prod_{j=1}^g\left(\rho_{2j-1}\rho_{2j} 
\rho_{2j-1}^{-1}\rho_{2j}^{-1}\right)=e^{2\pi i k/N}{\II}.  
\label{reprho}  
\end{equation}  
 
On the trivial bundle  
${\HH}\times {\CC}^N\rightarrow {\HH}$  
there is an action of $\Gamma$:  
$(z, v)\to(\gamma z, \rho(\gamma)v)$.  
This defines the quotient bundle  
\begin{equation}  
{\HH}\times {\CC}^N/\Gamma\rightarrow {\HH}/\Gamma\cong\Sigma.  
\label{trivialquot}  
\end{equation}  
Any admissible representation determines a holomorphic vector bundle  
$E_{\rho}\rightarrow \Sigma$ of rank $N$ and degree $k$.  
When $k=0$, $E_{\rho}$ is simply the  
quotient bundle (\ref{trivialquot}) of 
${\HH}\times {\CC}^N\rightarrow {\HH}$. The 
Narasimhan--Seshadri (NS) theorem \cite{NASE} 
now states that a holomorphic vector bundle $E$ over $\Sigma$ of rank $N$ and degree $k$  
is stable if and only if it is isomorphic to a bundle $E_{\rho}$, where $\rho$  
is an admissible representation of $\Gamma$.  
Moreover, the bundles $E_{\rho_1}$ and $E_{\rho_2}$  
are isomorphic if and only if the representations $\rho_1$  
and $\rho_2$ are equivalent.  
  
A differential--geometric approach to stability has been given by  
Donaldson \cite{DONALDSON}.  
Fix a Hermitean metric on $\Sigma$, 
for example the Poincar\'e metric, 
normalized so that the area of 
$\Sigma$ equals 1. Let us denote by $\omega$ its associated 
(1,1)--form. A holomorphic  
bundle $E$ is stable if and only if  
there exists on $E$ a metric connection $A_D$ with  
central curvature $F_D=-2\pi i \mu(E) \omega 
{\II}$; such a connection $A_D$ is unique. 
 
The unitary projective representations 
of $\Gamma$ we constructed above have a uniquely defined gauge field  
whose curvature is proportional to the volume form on $\Sigma$.  
With respect to the representation considered 
by NS, we note that NS introduced an elliptic point  
to produce the phase, while in our case the latter arises from the gauge length.  
Our construction is directly connected with Donaldson's approach as 
$F=iF_D$, where $F$ is the curvature 
(\ref{zummmolloo}). The main difference is that 
our operators are unitary 
differential operators on $L^2({\HH},{\CC}^N)$ instead of  
unitary matrices on ${\CC}^N$. 
This allowed us to obtain a non--trivial phase also in the Abelian case. 
 
It is however possible to understand the formal relation between  
our operators and those of NS. To see this we consider 
the adjoint representation 
of $\Gamma$ on ${\rm End}\,{\CC}^N$, 
\begin{equation}  
{\rm Ad}\,\rho (\gamma) Z =\rho (\gamma) Z \rho^{-1}(\gamma),  
\label{aggiunta}  
\end{equation}  
where $Z\in{\rm End}\, {\CC}^N$ is understood as an $N\times N$ matrix.  
Let us also consider the trivial bundle  
${\HH}\times{\rm End}\,{\CC}^N\rightarrow {\HH}$. 
The action of $\Gamma$ $(z,Z)\mapsto(\gamma z, {\rm Ad}\,\rho(\gamma) Z)$  
defines the quotient bundle  
\be 
{\HH}\times {\rm End}\, {\CC}^N/\Gamma\rightarrow {\HH}/  
\Gamma\cong\Sigma. 
\l{quotient2}\ee 
Then the idea is to consider a vector bundle $E'$ in the 
double scaling limit $N'\to\infty$, $k'\to-\infty$, with $\mu(E')=k'/N'$ fixed, that is 
$\mu(E')=b\mu(E)$. In this limit, fixing a basis in $L^2({\HH},\CC^N)$, the matrix elements  
of our operators can be identified with those of $\rho(\gamma)$. 
 
\vspace{.6cm} 
 
\noindent 
{\it 8. Noncommutative uniformization.} 
 
\noindent 
 Let us now introduce two copies of the upper half--plane, 
one with coordinates $z$ and $\bar z$, the other with coordinates $w$ and 
$\bar w$. While the coordinates $z$ and $\bar z$ are reserved to the 
operators ${\cal U}_k$ we introduced previously, we reserve $w$ and $\bar w$ 
to construct a new set of operators. We now introduce noncommutative 
coordinates expressed in terms of the covariant derivatives 
\be 
W=\partial_w+iA_w, \qquad \bar W=\partial_{\bar w}+iA_{\bar w}, 
\l{oeife}\ee 
with $A_w=A_{\bar w}=1/(2\,{\rm Im}\, w)$, so that $[W,\bar W]=iF_{w\bar w}$,  
where $F_{w\bar w}=i/[2({\rm Im}\, w)^2]$. 
Let us consider the following realization of the 
${\rm sl}_2({\RR})$ algebra: 
\begin{equation} 
\hat L_{-1}=-w,\qquad \hat L_0=-{1\over 2}(w\partial_w+\partial_w w), 
\qquad 
\hat L_1=-\partial_w w\partial_w. 
\label{sveglio}\end{equation} 
We then define the unitary operators 
\begin{equation} 
\hat T_k=e^{\lambda_{-1}^{(k)}(\hat L_{-1}+\bar{\hat L}_{-1})} 
e^{\lambda_0^{(k)}(\hat L_0+\bar{\hat L}_0)} 
e^{\lambda_{1}^{(k)}(\hat L_{1}+\bar{\hat L}_{1})}, 
\label{iVcappatree} 
\end{equation} 
where the $\lambda_n^{(k)}$ are as in (\ref{iVcappa}). 
Set ${\cal V}_k=\hat T_k{\cal U}_k$. Since 
the $\hat T_k$ satisfy (\ref{thetees}), it follows that the 
${\cal V}_k$ satisfy (\ref{abbiamox1nonabelianaduino}), times 
the $N\times N$ identity matrix, and 
\begin{equation} 
{\cal V}_k \partial_w {\cal V}_k^{-1}= 
\hat T_k \partial_w {\hat T_k}^{-1}={a_k\partial_w +b_k\over  
c_k\partial_w+d_k}. 
\label{dajjiee}\end{equation} 
Setting $W=G\partial_wG^{-1}$, {\it i.e.} $G=(w-\bar w)^2$, 
and using $Af(B)A^{-1}=f(ABA^{-1})$, we see that 
\begin{equation} 
{\cal V}_k W {\cal V}_k^{-1}= 
\hat T_k W {\hat T_k}^{-1} 
=G(\tilde w){\hat T_k} 
 \partial_w {\hat T_k}^{-1} G^{-1}(\tilde w), 
\label{poidhwd}\end{equation} 
where 
\begin{equation} 
\tilde w={\hat T_k} w {\hat T_k}^{-1}= 
-e^{-\lambda_0^{(k)}}+2\lambda_1^{(k)}(\hat L_0-\lambda_{-1}^{(k)}w) 
-\lambda_1^{(k)2}e^{\lambda_0^{(k)}}(\hat L_1+2\lambda_{-1}^{(k)} 
\hat L_0-{\lambda_{-1}^{(k)2}}w), 
\label{robbanova}\end{equation} 
and by (\ref{dajjiee}) 
\begin{equation} 
{\cal V}_k W {\cal V}_k^{-1}= 
\hat T_k W {\hat T_k}^{-1}={a_k\tilde W +b_k\over  
c_k\tilde W+d_k}, 
\label{dajjieedue}\end{equation} 
where 
\be 
\tilde W=\partial_w+G(\tilde w)[\partial_w G^{-1}(\tilde w)],  
\l{kjcns}\ee 
which differs from $W$ by the connection term. Eq.(\ref{dajjieedue}) 
can be seen as representing the noncommutative analog of uniformization. 
 
\vspace{.6cm} 
 
\noindent 
{\it 9. $C^\star$--algebra.} 
 
\noindent 
By a natural generalization of the $n$--dimensional 
noncommutative torus, one defines a noncommutative Riemann surface 
$\Sigma_\theta$ in $g>1$ to be an associative algebra with involution having 
unitary generators ${\cal U}_k$ obeying the relation (\ref{perfect}). 
Such an algebra is a $C^\star$--algebra, as it admits a faithful unitary  
representation on $L^2({\HH},{\CC}^N)$ whose image is norm--closed. 
Relation (\ref{perfect}) is also satisfied by the ${\cal V}_k$.  
However, while the ${\cal U}_k$ act on the commuting 
coordinates $z, \bar z$, the ${\cal V}_k$ act on the operators $W$ and $\bar W$. The latter, 
factorized by the action of the ${\cal V}_k$ in (\ref{dajjieedue}), can be pictorially identified 
with a sort of noncommutative  
coordinates on $\Sigma_\theta$.  
 
Each $\gamma\neq {\II}$ in 
$\Gamma$ can be uniquely expressed as a positive power of a primitive  
element $p\in\Gamma$, {\it primitive} meaning that $p$ is not a 
positive power of any other 
$p'\in\Gamma$ \cite{MCKEAN}. 
Let ${\cal V}_p$ be the representative of 
$p$. Any ${\cal V} \in C^\star$ can  
be written as 
\begin{equation} 
{\cal V}=\sum_{p\in\{prim\}}\sum_{n=0}^{\infty}c_n^{(p)}{\cal V}_p^n + c_0{\II}, 
\label{prim} 
\end{equation} 
for certain coefficients $c_n^{(p)}$, $c_0$. A trace  
can be defined as ${\rm tr}\,{\cal V}=c_0$. 
 
In the case of the torus one can connect the $C^\star$--algebras 
of $U(1)$ and $U(N)$. To see this one can use 't Hooft's 
clock and shift matrices $V_1$, $V_2$, which satisfy 
$V_1V_2=e^{2\pi i{M\over N}}V_2V_1$. 
The $U(N)$ $C^\star$--algebra  
is constructed in terms of the $V_k$ and of the unitary operators 
representing the $U(1)$ $C^\star$--algebra.  
Morita equivalence is an isomorphism 
between the two. 
In higher genus, the analog of the $V_k$ 
is the $U(N)$ representation $\rho(\gamma)$ 
considered above.  
One can obtain a $U(N)$ 
projective unitary differential representation of $\Gamma$ 
by taking ${\cal V}_k\rho(\gamma_k)$, 
with ${\cal V}_k$ Abelian. 
This nonabelian representation should be compared with 
the one obtained by the nonabelian ${\cal V}_k$ 
constructed above. In this framework it should be possible to 
understand a possible 
higher--genus analog of the Morita equivalence. 
 
The isomorphism of the 
$C^\star$--algebras is a direct consequence 
of an underlying equivalence between the $U(1)$ 
and $U(N)$ connection. The $z$--independence of the 
phase requires $F$ to be the identity matrix in the gauge indices. 
This in turn is deeply related to the uniqueness of the 
connection we found. The latter is 
related to the uniqueness of the NS connection. We conclude 
that Morita equivalence in higher genus 
is intimately related to the NS theorem. 
 
Our operators correspond to the $N\to\infty$ limit of 
projective unitary representations of $\Gamma$. These operators 
may be useful in studying the moduli space of M(atrix) string theory 
\cite{BONORA}. They also play a role in the $N\to\infty$ limit of QCD 
as considered in \cite{BOCHICCHIO}. 
 
Finally, let us note that an alternative proposal of noncommutative Riemann 
surfaces and $C^\star$--algebras has been considered in 
\cite{KlimekLesznewski}\cite{CHMM}. 
 
\vspace{1cm} 
 
{\bf Acknowledgments.} 
It is a pleasure to thank D. Bellisai, D. Bigatti, M. Bochicchio, 
G. Bonelli, L. Bonora, 
U. Bruzzo, R. Casalbuoni, G. Fiore, L. Griguolo, P.M. Ho, 
S. Kobayashi, I. Kra, G. Landi, K. Lechner, F. Lizzi, P.A. Marchetti, 
B. Maskit, F. R\u adulescu, D. Sorokin, W. Taylor, M. Tonin and 
R. Zucchini for comments and interesting discussions. G.B. is supported 
in part by a D.O.E. cooperative agreement DE-FC02-94ER40818 and by an INFN ``Bruno 
Rossi" Fellowship. J.M.I. is supported by an INFN fellowship. J.M.I., M.M. and 
P.P. are partially supported by the European Commission TMR program 
ERBFMRX-CT96-0045.


\begin{thebibliography}{99} 
 
\bibitem{MATRIX}  
T. Banks, W. Fischler, S. Shenker and L. Susskind, Phys. Rev. {\bf D55} 
(1997) 5112; 
for reviews see  
D. Bigatti and L. Susskind, hep-th/9712072; T. Banks, hep-th/9911068; 
W. Taylor, hep-th/9801182. 
\bibitem{JAP} 
N. Ishibashi, H. Kawai, Y. Kitazawa and A. Tsuchiya, Nucl. Phys. {\bf B498} 
(1997) 467. 
\bibitem{SUSS} 
L. Susskind, hep-th/9704080. 
\bibitem{REVTAYLOR} 
W. Taylor, hep-th/0002016. 
\bibitem{TAYLOR}  
W. Taylor, Phys. Lett. {\bf B394} (1997) 283; O. Ganor, S. Ramgoolam and W. Taylor, 
Nucl. Phys. {\bf B492} (1997) 191. 
\bibitem{CDS}  
A. Connes, M. Douglas and A. Schwarz, JHEP {\bf 02} 
(1998) 003. 
\bibitem{HULL}  
M. Douglas and C. Hull, JHEP {\bf 02} 
(1998) 008; C. Hull, JHEP {\bf 10} (1998) 011. 
\bibitem{BIGATTI}  
D. Bigatti, Phys. Lett. {\bf B451} (1999) 324; hep-th/9802129; 
Ph.D. Thesis. 
\bibitem{CASALBUONI} 
R. Casalbuoni, Phys. Lett. {\bf B431} (1998) 69. 
\bibitem{CONNESETAL}  
A. Connes, {\it Noncommutative Geometry}, Academic Press, London (1994); 
hep-th/0003006; 
G. Landi, {\it An Introduction to Noncommutative Spaces and their Geometry},  
Lecture 
Notes in Physics: Monographs, m51, Springer--Verlag, Berlin--Heidelberg (1997); 
M. Douglas, hep-th/9901146. 
\bibitem{DOUGLAS} 
M. Douglas, H. Ooguri and S. Shenker, Phys. Lett. {\bf B402} 
(1997) 36; 
M. Douglas and H. Ooguri, Phys. Lett. {\bf B425} (1998) 71. 
\bibitem{SCHWARZ} 
N. Nekrasov and A. Schwarz, Comm. Math. Phys. {\bf 198} 
(1998) 689; 
A. Astashkevich, N. Nekrasov and A. Schwarz, hep-th/9810147;  
A. Konechny and A. Schwarz, Nucl. Phys. {\bf B550} 
(1999) 561; Phys. Lett. {\bf B453} (1999) 23. 
\bibitem{VERLINDE}  
C. Hofman and E. Verlinde, JHEP 
{\bf 12} (1998) 010; Nucl. Phys. {\bf B547} 
(1999) 157. 
\bibitem{ZUMINO}  
B. Moriaru and B. Zumino, hep-th/9807198. 
\bibitem{LLS}  
G. Landi, F. Lizzi and R. Szabo, Comm. Math. Phys. {\bf 206} 
(1999) 603. 
\bibitem{HO} 
P.-M. Ho, Y.-Y. Wu and Y.-S. Wu, Phys. Rev. {\bf D58} (1998) 
026006; 
P.-M. Ho and Y.-S. Wu, Phys. Rev. {\bf D58} (1998) 066003; 
Phys. Rev. {\bf D60} (1999) 026002; 
P.-M. Ho, Phys. Lett. {\bf B434} (1998) 41. 
\bibitem{IRAN} 
F. Ardalan, H. Arfaei and M.M. Sheikh-Jabbari, 
JHEP {\bf 02} (1999) 016. 
\bibitem{SWn} 
N. Seiberg and E. Witten, JHEP {\bf 09} 
(1999) 032. 
\bibitem{SchCorSch}  
V. Schomerus, JHEP {\bf 06} (1999) 030; 
L. Cornalba and R. Schiappa, hep-th/9907211; 
L. Cornalba, hep-th/9909081, hep-th/9912293; 
Y. Imamura, JHEP {\bf 01} (2000) 039; 
H. Hata and S. Moriyama, JHEP {\bf 02} (2000) 011; 
J. Madore, S. Schraml, P. Schupp and J. Wess, hep-th/0001203; 
I.Ya. Aref'eva, D.M. Belov and A.S. Koshelev, hep-th/0001215; 
M.B. Green and M. Gutperle, hep-th/0002011; 
P. Bouwknegt and V. Mathai, hep-th/0002023; 
A. Matusis, L. Susskind and N. Toumbas, hep-th/0002075; 
K. Hashimoto and T. Hirayama, hep-th/0002090; 
A. Yu. Alekseev and A.G. Bytsko, hep-th/0002101; 
T. Asakawa and I. Kishimoto, hep-th/0002138; 
J. Ambjorn, Y.M. Makeenko, J. Nishimura, R.J. Szabo, 
hep-th/0002158; P.-M. Ho, hep-th/0003012. 
\bibitem{NOI} 
G. Bertoldi, J.M. Isidro, M. Matone and P. Pasti, to appear. 
\bibitem{THOMPSON} G. Thompson, hep-th/9305120. 
\bibitem{CHMM} A. Carey, K. Hannabuss, V. Mathai and P. McCann, 
Comm. Math. Phys. {\bf 190} (1998) 629; 
A. Carey, K. Hannabuss and V. Mathai, Lett. Math. Phys. {\bf 47} 
(1999) 215. 
\bibitem{RADULESCU}  
F. R\u adulescu, {\it Memoirs of the American Mathematical Society} {\bf 630}  
(1998) vol. 133; math/9912033. 
\bibitem{KOBAYASHI}  
S. Kobayashi, {\it Differential Geometry of Complex Vector Bundles}, Iwanami 
Shoten Publishers and Princeton University Press, Princeton (1987); 
R. Gunning, {\it Lectures on Vector Bundles over Riemann Surfaces}, 
Princeton Univ. Notes 1967; 
 J. Fay, 
{\it Memoirs of the American Mathematical Society} {\bf 464}  
(1992) vol. 96. 
\bibitem{AtiyahSenGuptaFine} M. Atiyah and R. Bott, {\it Phil. Trans. Roy. 
Soc. London} {\bf A 308} (1982) 524; P.G. Zograf and L.A. Takhtajan, 
Math. USSR Izvestiya {\bf 35} (1990) 83; 
D. Fine, Comm. Math. Phys. {\bf 140} (1991) 321; 
F. Ferrari, Helv. Phys. Acta {\bf 67} (1994) 702; 
A. Sengupta, {\it Memoirs of the American Mathematical Society} {\bf 600}  
(1997) vol. 126. 
\bibitem{NASE} 
M. Narasimhan and C. Seshadri, {\it Ann. Math.} {\bf 82} (1965) 540. 
\bibitem{DONALDSON}  
S. Donaldson, {\it Journ. Diff. Geom.} {\bf 18} (1983) 269. 
\bibitem{MCKEAN} 
H. McKean, {\it Comm. Pure Applied Math.} {\bf 25} (1972) 225. 
\bibitem{BONORA} 
G. Bonelli, L. Bonora and F. Nesti, Nucl. Phys. {\bf B538} (1999) 100; 
G. Bonelli, L. Bonora, F. Nesti and A. Tomasiello, Nucl. Phys. {\bf B554} (1999) 103. 
\bibitem{BOCHICCHIO} M. Bochicchio,  JHEP {\bf 01} (1999) 005; 
{\bf 01} (1999) 006; hep-th/9904200; hep-th/9904201. 
\bibitem{KlimekLesznewski} S. Klimek and A. Lesznewski, Comm. Math. Phys. 
{\bf 146} (1992) 105; Lett. Math. Phys. {\bf 24} (1992) 125. 
 
\end{thebibliography}
\end{document}